%
%
%

\def\oddpage{{\tt preliminary draft \hfil \jobname \hfil \today}}
\def\evenpage{{\tt \today \hfil\jobname \hfil preliminary draft}}
\def\titlepage{{\tt preliminary draft \hfill \jobname}}
\def\draft{\baselineskip = 16pt plus 2pt minus 1pt
   \hsize = 17.0 truecm \vsize = 24.7 truecm
   \hoffset=-4 truemm   
    \overfullrule = 5pt
    \headline{\ifnum\count0>1\ifodd\count0\oddpage\else%
    \evenpage\fi\else\titlepage\fi}
}
\overfullrule=0pt             
\hoffset=-3 true mm           
\voffset=9.5 true mm

%
%
%
%

\newif\iflabels

\newif\ifreference


\newcount\secnum    \global\secnum=0
\newcount\subsecnum \global\subsecnum=0
\newcount\eqnum     \global\eqnum=0
\newcount\citenum   \global\citenum=0
\newcount\fignum    \global\fignum=0
\newcount\tablenum  \global\tablenum=0
\newcount\remarknum    \global\remarknum=0

\newif\ifndouble
\def\doublenumbers{\ndoubletrue\gdef\therunningsection{\the\secnum}}

\def\ifundefined#1{\expandafter\ifx\csname#1\endcsname\relax}
\def\strip#1>{}


\def\cref#1{\ifundefined{@c@#1}\immediate\write16{ --> \string\cref{#1}
    not defined !!!}
    \expandafter\xdef\csname@c@#1\endcsname{??}\fi\csname@c@#1\endcsname}

\def\eqref#1{\ifundefined{@eq@#1}\immediate\write16{ --> \string\eqref{#1}
    not defined !!!}
    \expandafter\xdef\csname@eq@#1\endcsname{??}\fi\csname@eq@#1\endcsname}

\def\sref#1{\ifundefined{@s@#1}\immediate\write16{ --> \string\sref{#1}
    not defined !!!}
    \expandafter\xdef\csname@s@#1\endcsname{??}\fi\csname@s@#1\endcsname}

\def\figref#1{\ifundefined{@f@#1}\immediate\write16{ -->
    \string\figref{#1} not defined !!!}
    \expandafter\xdef\csname@f@#1\endcsname{??}\fi\csname@f@#1\endcsname}

\def\tableref#1{\ifundefined{@f@#1}\immediate\write16{ -->
    \string\tableref{#1} not defined !!!}
    \expandafter\xdef\csname@f@#1\endcsname{??}\fi\csname@f@#1\endcsname}

\newdimen\beforesecskip  \beforesecskip=\baselineskip
\newdimen\aftersecskip   \aftersecskip=0pt

\font\sectionfont = cmbx10 at 11 pt

\def\Section#1#2{            
    \global\advance\secnum by 1\ifndouble\global\eqnum=0\fi
    \global\subsecnum=0
    \xdef\therunningsection{\the\secnum}
    \def\usenthrow{1}\ifundefined{@s@#1}\def\usenthrow{2}\fi
    \expandafter\ifx\csname@s@#1\endcsname\therunningsection\def\usenthrow{2}\fi
    \ifodd\usenthrow\immediate\write16
      { --> Possible reference error in \string\sref{#1} }\fi
    \expandafter\xdef\csname@s@#1\endcsname{\therunningsection}
    \immediate\write16{\therunningsection. #2}
    \goodbreak\vskip\beforesecskip\noindent%
    \iflabels
          \llap{\tt #1\quad}%
    \fi
    {\sectionfont\the\secnum.\enspace#2}\par\nobreak\noindent\ignorespaces}

\def\subsection#1{\global\advance\subsecnum by 1%
    \xdef\therunningsubsection{\the\subsecnum}%
    \medbreak\smallskip
    \noindent{\sl\the\secnum.\the\subsecnum.\enspace #1}%
    \nobreak\noindent}

\def\section#1#2{\Section{#2}{#1}\par\noindent\ignorespaces}

\def\asection#1{\immediate\write16{#1}%
    \goodbreak\vskip\beforesecskip
    \noindent{\sectionfont#1}\par\nobreak\noindent\ignorespaces}


\def\eqlabel#1{\global\advance\eqnum by 1
    \ifndouble\xdef\anumber{\therunningsection.\the\eqnum}
       \else\xdef\anumber{\the\eqnum}\fi
    \def\usenthrow{1}\ifundefined{@eq@#1}\def\usenthrow{2}\fi
    \expandafter\ifx\csname@eq@#1\endcsname\anumber\def\usenthrow{2}\fi
    \ifodd\usenthrow\immediate\write16
       { --> Possible reference error in \string\eqref{#1} }\fi
    \expandafter\xdef\csname@eq@#1\endcsname{\anumber}
    \ifndouble
       \def\usenthrow{\expandafter\strip\meaning\therunningsection.\the\eqnum}
       \else\def\usenthrow{\the\eqnum}\fi
}

\def\autoeqno#1{\eqlabel{#1}\eqno(\csname@eq@#1\endcsname)
    \iflabels \rlap{\quad\tt #1} \fi
}
\def\autoleqno#1{\eqlabel{#1}\leqno(\csname@eq@#1\endcsname)
    \iflabels \llap{\tt #1 \qquad} \fi
}

\def\therefs{}
\def\bibitem#1#2\par{
    \global\advance\citenum by 1
    \xdef\citation{\the\citenum}
    \def\usenthrow{1}\ifundefined{@c@#1}\def\usenthrow{2}\fi
    \expandafter\ifx\csname@c@#1\endcsname\citation\def\usenthrow{2}\fi
    \ifodd\usenthrow\immediate\write16
      { --> Possible reference error in \string\cref{#1} }\fi
    \expandafter\xdef\csname@c@#1\endcsname{\citation}
\iflabels
     \ifnum\citenum = 1\global\xdef\therefs{\par\noindent\llap{\tt#1\qquad}%
          \ignorespaces#2\par}
     \else 
          \global\xdef\oldrefs{\therefs}
          \global\xdef\therefs{\oldrefs\par\noindent\llap{\tt#1\qquad}%
          \ignorespaces#2\par}
     \fi
\else
     \ifnum\citenum = 1\global\xdef\therefs{\item{[\citation]} #2\par }
     \else 
          \global\xdef\oldrefs{\therefs}
          \global\xdef\therefs{\oldrefs\item{[\citation]} #2\par }%
     \fi
\fi
}


\def\cite#1{\hbox{[\cref{#1}]}}

\newcount\refcount
\refcount=1
\def\listrefs{\frenchspacing
    \asection{References}\par
    \iflabels
          {
          \everypar{\hang\textindent{[\the\refcount]}
          \global\advance\refcount by 1\relax}\therefs
          }
    \else
          \therefs
    \fi
    \nonfrenchspacing}



\newdimen\captionwidth
\captionwidth = \hsize
\advance\captionwidth by -2\parindent
\newbox\captionbox

\def\figure#1#2#3{
    \global\advance\fignum by 1
    \xdef\afigure{\the\fignum}
    \def\usenthrow{1}\ifundefined{@f@#1}\def\usenthrow{2}\fi
    \expandafter\ifx\csname@f@#1\endcsname\afigure\def\usenthrow{2}\fi
    \ifodd\usenthrow\immediate\write16
      { --> Possible reference error in \string\figref{#1} }\fi
    \expandafter\xdef\csname@f@#1\endcsname{\afigure}
    \ifnum\fignum = 1\global\xdef\thefigs{\item{Fig.\ \afigure.} #2\ }
    \else%
    \global\xdef\oldfigs{\thefigs}%
    \global\xdef\thefigs{\oldfigs\item{Fig.\ \afigure:} #2\ }%
    \fi%
     \goodbreak\midinsert
     \ifx\epsfbox\undefined
          \immediate\write16{ Fig. \afigure: ignored }
          \noindent\hrule\par\vskip 1cm \noindent\hrule
     \else\immediate\write16{ Fig. \afigure.}
          \center{#3}          
     \fi
     \smallskip
     \setbox\captionbox=\hbox{Figure \afigure: \ignorespaces#2}
     \iflabels
          \ifdim \wd\captionbox < \captionwidth
               \noindent\llap{\tt#1\quad}\centerline{Figure \afigure: \ignorespaces#2}
          \else
               \noindent
               \llap{\tt#1\quad}{\narrower\noindent Figure \afigure: \ignorespaces#2\par}
          \fi
     \else
          \ifdim \wd\captionbox < \captionwidth \centerline{Figure \afigure: \ignorespaces#2}
          \else {\narrower\noindent Figure \afigure: \ignorespaces#2\par}
          \fi
     \fi
     \endinsert
}


\def\table#1#2#3{
    \global\advance\tablenum by 1
    \xdef\atable{\the\tablenum}
    \def\usenthrow{1}\ifundefined{@f@#1}\def\usenthrow{2}\fi
    \expandafter\ifx\csname@f@#1\endcsname\atable\def\usenthrow{2}\fi
    \ifodd\usenthrow\immediate\write16
      { --> Possible reference error in \string\tableref{#1} }\fi
    \expandafter\xdef\csname@f@#1\endcsname{\atable}
    \ifnum\tablenum = 1\global\xdef\thetables{\item{Table \atable.} #2\ }
    \else%
    \global\xdef\oldtables{\thetables}%
    \global\xdef\thetables{\oldtables\item{Table \atable.} #2\ }%
    \fi%
     \goodbreak\midinsert
     \immediate\write16{ Table \atable.}
     \setbox\captionbox=\hbox{Table \atable. \ignorespaces#2}
     \iflabels
          \ifdim \wd\captionbox < \captionwidth
               \noindent\llap{\tt#1\quad}\centerline{Table \atable. \ignorespaces#2}
          \else
               \noindent
               \llap{\tt#1\quad}{\narrower\noindent Table \atable. \ignorespaces#2\par}
          \fi
     \else
          \ifdim \wd\captionbox < \captionwidth \centerline{Table \atable. \ignorespaces#2}
          \else {\narrower\noindent Table \atable. \ignorespaces#2\par}
          \fi
     \fi
     \medskip
     \let\\=\cr
     \centerline{\vbox{\offinterlineskip\halign{\strut\ignorespaces#3}}}
     \medskip
     \endinsert
}


\def\hline{\noalign{\hrule}}
\def\today{\ifcase\month\or January\or February\or
   March\or April\or May\or June\or July\or August\or September\or
   October\or November\or December\fi
   \space\number\day, \number\year}
\def\frac#1#2{{#1\over#2}}

\def\text#1{{\rm #1}}
\def\degs{\ifmmode {}^\circ \else ${}^\circ$ \fi} 
\def\[{\begingroup$$\let\\=\cr}
\def\]{$$\endgroup\ignorespaces}
\def\\{\hfil\break}
\def\){\hfill\break}

\def\roughly#1{\mathrel{\raise.3ex\hbox{$#1$\kern-.75em\lower1ex%
\hbox{$\sim$}}}}

\let\thanks=\footnote
\newcount\fnotenum\fnotenum=0
\def\footnote#1{\advance\fnotenum by 1 \thanks{$^{\the\fnotenum}$}{#1}}

\newcount\itemnum
\let\Item=\item
\def\beginenumerate{\itemnum=0\relax \par\begingroup\nobreak
\def\item{\advance\itemnum by 1 \Item{\the\itemnum.}}}



\def\titleparagraphs{\interlinepenalty=9999
     \leftskip=0.03\hsize plus 0.22\hsize minus 0.03\hsize
     \rightskip=\leftskip \parfillskip=0pt
     \hyphenpenalty=9000 \exhyphenpenalty=9000
     \tolerance=9999 \pretolerance=9000
     \spaceskip=0.333em \xspaceskip=0.5em }
\def\center#1{\par{
     \def\\{\break} \titleparagraphs \noindent #1\par}}

\def\preprint#1{\par\rightline{\tt #1}\bigskip}

\def\remark{\global\advance\remarknum by 1
   \smallskip{\it Remark \the\remarknum.}\enspace\ignorespaces}

\font\titlefont = cmbx10 at 12 pt
\def\title#1{\center{\baselineskip=14pt\titlefont\ignorespaces#1}}
\def\author#1{\bigskip\center{\ignorespaces\bf#1}}
\def\address#1{\smallskip\center{\ignorespaces#1}}
\def\date#1{\medskip\centerline{#1}}


\input epsf
\let\label=\autoeqno
\let\ref=\eqref

\def\bfig#1{\epsfysize=0.355\vsize \epsfbox{#1.eps}}
\def\textbf#1{{\bf#1}}
\def\texttt#1{{\tt#1}}
\def\mbox#1{\hbox{#1}}

\hyphenation{an-a-lyse be-hav-iour}
\def\Ne{N_{\rm e}}

\bibitem{we1}
   M. Yu. Zotov, G. V. Kulikov, Izvestiya RAN, ser. fiz., \textbf{68}
   (2004)~1602 (in Russian).

\bibitem{we2}
   G. V. Kulikov, M. Yu.\ Zotov, Preprint \texttt{astro-ph/0407138}.

\bibitem{knee}
   G. V. Kulikov, G. B. Khristiansen, ZhETF \textbf{35} (1958)~635.

\bibitem{Al}
   D. E. Alexandreas, D. Berley, S. Biller et al.,
   Astrophys.~J. \textbf{383} (1991) L53.

\bibitem{Milagrito}
   K. Wang, R. Atkins, W. Benbow et al., Astrophys.~J. \textbf{558} (2001) 477.

\bibitem{Milagro}
   P. M. Saz Parkinson, Preprint \texttt{astro-ph/0503244}.

\bibitem{HEGRA}
   F. Aharonian, A. Akhperjanian, J.~A.~Barrio et al., Astron. Astrophys.
   \textbf{390} (2002) 39.

\bibitem{KASCADE}
   T. Antoni, W. D. Apel, A. F. Badea et al., Astrophys.~J. \textbf{608}
   (2004) 865; \texttt{astro-ph/0402656}.

\bibitem{Fomin}
   Yu.~A. Fomin et al., Proc.\ 26th ICRC, Salt Lake City, 1999,
   v.~1, p.~286.

\bibitem{AGASA}
   N. Hayashida, K. Honda, N. Inoue et al., Astrophys.~J. \textbf{522}
   (1999)~225.

\bibitem{Green}
   D. A. Green, 2006, `A Catalogue of Galactic Supernova Remnants (2006
   April version),' Astrophysic Group, Cavendish Laboratory, Cambridge,
   United Kingdom\\
   (available at \mbox{\texttt{http://www.mrao.cam.ac.uk/surveys/snrs/}}).

\bibitem{ATNF}
   The ATNF Pulsar Database,
   \mbox{\texttt{http://www.atnf.csiro.au/research/pulsar/psrcat/}};\\
   R.~N.~Manchester, G.~B.~Hobbs, A.~Teoh, M.~Hobbs, Astrophys.~J. \textbf{129}
   (2005)~1993; \texttt{astro-ph/0412641}.

\bibitem{Simbad}
   The SIMBAD database, \mbox{\texttt{http://simbad.u-strasbg.fr/Simbad}}

\bibitem{Octave}
   J. W. Eaton, ``GNU Octave: A High-Level Interactive 	
   Language for Numerical Computations.'' Edition~3 for version 2.0.13, 1997;
   \mbox{\texttt{http://www.octave.org/}}

\preprint{astro-ph/0610944}

\title{Search for Sources of Cosmic Rays in the Region of the Knee.~II}

\author{G. V. Kulikov, M. Yu.\ Zotov}
\address{
   D. V. Skobeltsyn Institute of Nuclear Physics\\
   Moscow State University, Moscow 119992, Russia\\
   \tt $\{$kulikov,zotov$\}$@eas.sinp.msu.ru}

\date{October 31, 2006}

\bigskip
\centerline{\bf Abstract}
{\narrower\noindent
New results of an analysis of arrival directions of extensive air showers
registered with the EAS--1000 Prototype array from August 1997 till
February 1999 are presented. The method of Alexandreas et al., which has
been used for analysis of data registered with CYGNUS, Milagrito, HEGRA
AIROBICC, KASCADE and a number of other experiments, is employed. The
existence of zones of excessive flux of cosmic rays with energies in the
region of the knee is confirmed, as well as closeness of the zones to
coordinates of possible astrophysical cosmic ray sources.

}

\section{Introduction}{sec:intro}
   We have already studied arrival directions of extensive air showers
   (EAS) with energies of primary particles of the order of
   $10^{14}$--$10^{15}$~eV registered with the EAS--1000 Prototype
   array~[\cref{we1},~\cref{we2}].
   In these works, the analysis was performed with the help of a special
   procedure of filtering experimental data which allowed us to obtain a
   set of data such that at some scale, the showers were distributed
   uniformly wrt.\ sidereal time and the azimuth angle.
   In this method, filtering of data was done in a (pseudo-)random way.
   Thus, in what follows, we shall call this procedure as a method of
   random filtering~(MRF).
   A detailed description of the MRF can be found in~[\cref{we1},~\cref{we2}].

   The MRF applied to our data set resulted in the selection of 37~zones
   of excessive flux (ZEFs) of EAS.
   Location of the majority of the ZEFs was found to be close to
   the coordinates of possible astrophysical sources of cosmic rays (CRs)
   with energies in the region of the knee in the energy spectrum
   at around 3~PeV~\cite{knee}.
   The fact that application of the MRF to the available data set
   led to exclusion of about~17\% of the initial number of EAS with
   known arrival directions, motivated us to perform a similar
   analysis basing on the method of Alexandreas et al.~\cite{Al}.
   This method utilizes the full amount of available data.
   It has been successfully employed earlier in a number of investigations,
   see, e.g., reports on the analysis of data obtained with the
   CYGNUS~\cite{Al}, Milagrito~\cite{Milagrito}, Milagro~\cite{Milagro},
   HEGRA AIROBICC~\cite{HEGRA}, and KASCADE~\cite{KASCADE} experiments.
   Below we present the results of our investigation.

\section{Experimental Data}{sec:data}
   The data set under consideration includes 1,668,489 EAS registered
   during 203 days of operation of the EAS--1000 Prototype Array in the
   period from August~30, 1997, till February~1, 1999.
   The array consisted of eight detectors 
   located in the central part of the EAS MSU array along longer sides 
   of the $64\,{\rm m}\times22\,{\rm m}$ rectangle~\cite{Fomin}.
   The geographical coordinates of the array were
   $37^\circ32.5'$E, $55^\circ41.9'$N.

   Arrival directions were determined for 1,366,010 EAS.
   A number~$\Ne$ of charged particles (electrons) was found for 826,921
   EAS.
   The mean value $\bar\Ne=1.20\times10^5$ corresponds to the energy
   of a primary proton $E\approx10^{15}$~eV.
   There are only 11262 EAS with $\Ne>10^6$ but they give a noticeable
   contribution in the value of~$\bar\Ne$.
   With the above energetic EAS excluded, $\bar\Ne=7.65\times10^4$
   and the median value equals $3.62\times10^4$.
   Thus, the majority of showers in the data set under consideration
   had primary particles with energies slightly below the knee.

\section{Method of Investigation}{sec:method}
   The main idea of the method of Alexandreas et al.~\cite{Al}
   is the following.
   A search for ZEFs is based on the comparison of real experimental data
   set with a ``background'' set that would be registered with the same
   installation and during the same period of observation in the
   assumption of a uniform distribution of arrival directions of EAS over
   the celestial sphere.
   The background data set is obtained by a simple procedure of
   randomization.
   First, for every shower with an arrival direction $(\theta,\varphi)$
   an arrival time~$t'$ of another
   shower is chosen randomly from the rest of the experimental data set.
   Next, arrival directions of all showers in the equatorial coordinates
   $(\alpha,\delta)$ are calculated basing on the respective triplets
   $(t',\theta,\varphi)$.
   In the resulting set, the distribution of EAS wrt.~$\delta$ is exactly
   the same as in the original data set.

   Randomization is performed multiple times in order to avoid
   a dependency of the distribution wrt.~$\alpha$ on the concrete
   choice of~$t'$.
   After this, one creates an averaged ``background'' map by
   calculating the mean number of EAS located in cells of a
   fixed size $\Delta\alpha\times\Delta\delta$ in all obtained maps.
   It is believed that the background map has most of the
   properties of an isotropic background~\cite{KASCADE}.
   A search for ZEFs is performed by comparing the number of EAS
   in a region in the original (``real'') experimental map
   with the same region in the background map.
   The measure of deviation is the significance
   $S=(N_{\rm real} - N_{\rm bg})/\sqrt{N_{\rm bg}}$, where
   $N_{\rm real}$ and $N_{\rm bg}$ are the numbers of EAS in
   the same region of the ``real'' experimental map and the
   background maps respectively.

   For the purposes of this investigation, the background map
   was obtained by averaging the number of showers binned in
   $\Delta\alpha\times\Delta\delta=1^\circ\times1^\circ$ cells.
   An analysis of the background maps revealed that two
   maps obtained consecutively differ by at most~1\% of the number
   of EAS in the respective cells after around~50 cycles of
   randomization, and by~0.5\% after~90 cycles.
   Thus we chose to perform 100 cycles of randomization
   to obtain the final background map.

   To look for ZEFs, we followed the same procedure as we used
   in~[\cref{we1}, \cref{we2}].  Namely, the field of observation was
   divided into strips of width $\Delta\delta=3^\circ\dots30^\circ$,
   with the boundaries of the strips of equal width shifted by~$1^\circ$
   wrt.\ each other.
   Each strip was then divided into cells of equal width~$\Delta\alpha$.
   In the main part of the investigation, for any fixed width~$\Delta\delta$
   of a strip, $\Delta\alpha$ was chosen to be equal to a rounded value
   (expressed in degrees) of $\Delta\delta/\cos\bar{\delta}$, where~$\bar{\delta}$
   is the mean value of the declination for the strip.
   This guarantees that cells with the same~$\Delta\delta$ but located
   at different declination, have an approximately equal area.
   Similar to~[\cref{we1}, \cref{we2}], such cells will be called `regular'.
   It is important to take into account that neither the method
   of Alexandreas et al., nor the method of random filtering put
   a restriction on the way one chooses regions to be compared.
   One may perform a search for ZEFs with an arbitrary relation
   of~$\Delta\alpha$ and~$\Delta\delta$, as well as ZEFs of an
   arbitrary shape.

\section{Main Results}{sec:main}
   Application of the method of Alexandreas et al. to the available data
   set resulted in the discovery of~561 regular cells with $S>3.0$.
   For 364 of these cells, significance $S>3.1$.
   For the sake of convenience, the cells were joined into 27 zones
   such that each zone is a connected domain.
   In order to perform an accurate comparison of the results of the two
   methods, we added five zones made of ``irregular'' cells, i.e., cells
   with an arbitrary relation of~$\Delta\alpha$ and~$\Delta\delta$.
   The resulting ZEFs are shown in Fig.~1 in red.
   The most of partially overlapping cells of excessive flux
   are not shown for the sake of visual clarity.
   Their joint exterior boundaries are shown instead.
   Blue lines show the boundaries of the ZEFs presented
   in~[\cref{we1}, \cref{we2}].  Recall that all of these 37 ZEFs
   but zones No.~4, 5, 6, 20, and~21, and bigger cells in ZEFs
   No.~3, 27, 29, 33, and~36 are composed of regular cells.
   A number of the zones found by the MRF coincide with those
   found by the method of Alexandreas et al.\ and thus their
   boundaries are not visible in the figure.

   \figure{fig:zefs}{Red lines show the boundaries of ZEFs found by the
   method of Alexandreas et al. Blue lines show the boundaries of ZEFs
   found by the method of random filtering. Numbers in the field of the
   figure show the numbers of ZEFs assigned to them in~[\cref{we1},
   \cref{we2}]. The (magenta) arcs show the Galactic plane. The $\cap$-like
   curve shows the Supergalactic plane.}{\bfig{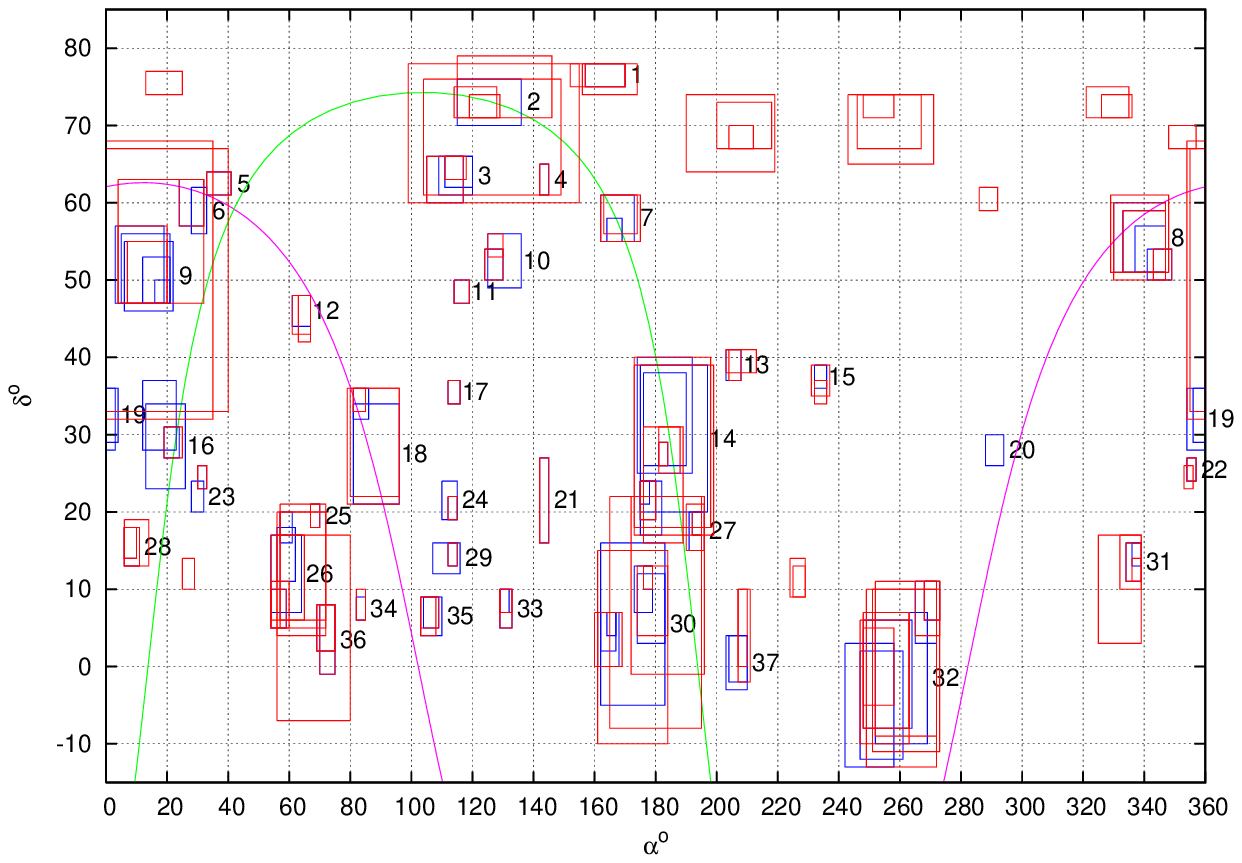}}

   As is clear from Fig.~1, the results obtained by two different methods
   are quite close, though the MRF seems to be more selective in the
   sense that the total area of ZEFs found by this method is less than
   that of the zones found by the method of Alexandreas et al.\
   (``A-zones''). In sum, 20 ZEFs found by the MRF lie {\it inside\/}
   A-zones. These are zones No.~1--9, 12, 14, 15, 18, 22, 25--28, 31,
   and~36. Zone No.~30 also lies inside the corresponding A-zone
   excluding a small area
   $\Delta\alpha\times\Delta\delta=3^\circ\times1^\circ$. A major part
   of ZEF No.~32 also lies within the corresponding A-zone. Zones No.~11,
   17, and~21 exactly coincide with A-zones, while zones No.~13 and
   33--35 coincide with A-zones up to~$1^\circ$. In fact, there are
   multiple coincidences with deviation of at most~$1^\circ$ of separate
   cells found my the two methods  in all cases when a ``composite'' ZEF
   found by the MRF is embedded into a zone found by the method of
   Alexandreas et~al.

   Let us mention the existence of pairs for the central cell in ZEF
   No.~16, which has the M33 galaxy located inside, and for ZEF No.~7,
   which contains the famous C2 triplet of ultra-high energy cosmic rays,
   registered with the AGASA array~\cite{AGASA}. Recall that the
   observation of the latter coincidence was one of the motivations for
   the analysis of coordinates of extra-galactic cosmic ray sources
   performed in~\cite{we2}.

   An embedding of A-zones into the ZEFs found by the MRF can be seen in
   a few cases. Namely, ZEF No.~10, 23, 24, and~29 contain the
   corresponding A-zones inside. Only three of 37 ZEFs found by the MRF
   do not have close counterparts among A-zones. These are the irregular
   ZEF No.~20, the irregular (exterior) cell of ZEF No.~29, and ZEF
   No.~37.  In the latter case, there are a number of irregular A-zones
   that intersect the ZEF.  They are shown in the figure.

   Finally, notice seven A-zones (made of regular cells) that do not
   have any close counterparts among the ZEFs found by the MRF.
   These are a zone located around $\alpha=20^\circ,\,\delta=75^\circ$,
   two big A-zones in the vicinity of $\delta=70^\circ$,
   a zone near $\alpha=330^\circ,\,\delta=73^\circ$,
   a zone around $\alpha=290^\circ,\,\delta=60^\circ$,
   and two zones near $\delta=10^\circ$ ($\alpha=25\dots29^\circ$ and
   $\alpha=224\dots229^\circ$ respectively).

\section{Discussion}{sec:disc}
   In our opinion, there is a qualitative agreement between the results
   of the analyses of arrival directions of cosmic rays registered with
   the EAS--1000 Prototype Array performed by two different methods.
   The agreement confirms one of the main conclusions
   of~[\cref{we1}, \cref{we2}] that there are zones of excessive flux
   of EAS in the available data set.
   We find it remarkable that mutual coincidences or embeddings are
   observed for all zones located in the vicinity of the Galactic
   and Supergalactic planes.

\figure{fig:objects}{Zones of excessive flux found by the
   method of Alexandreas et~al.~\cite{Al} that have counterparts
   among the ZEFs found in~[\cref{we1}, \cref{we2}].
   Different symbols show coordinates of some astrophysical objects
   that have coordinates within angular distance $\le3^\circ$
   from the ZEFs.  These are galactic supernova remnants~$(\bullet)$,
   pulsars (filled blue triangles), active galactic nuclei and 
   interacting galaxies at red shifts $z\le0.01$~($\triangle$).
   Also shown are the coordinates of the C2 triplet registered
   with the AGASA array~($*$)~\cite{AGASA} and the M33
   galaxy (the red box).}{\bfig{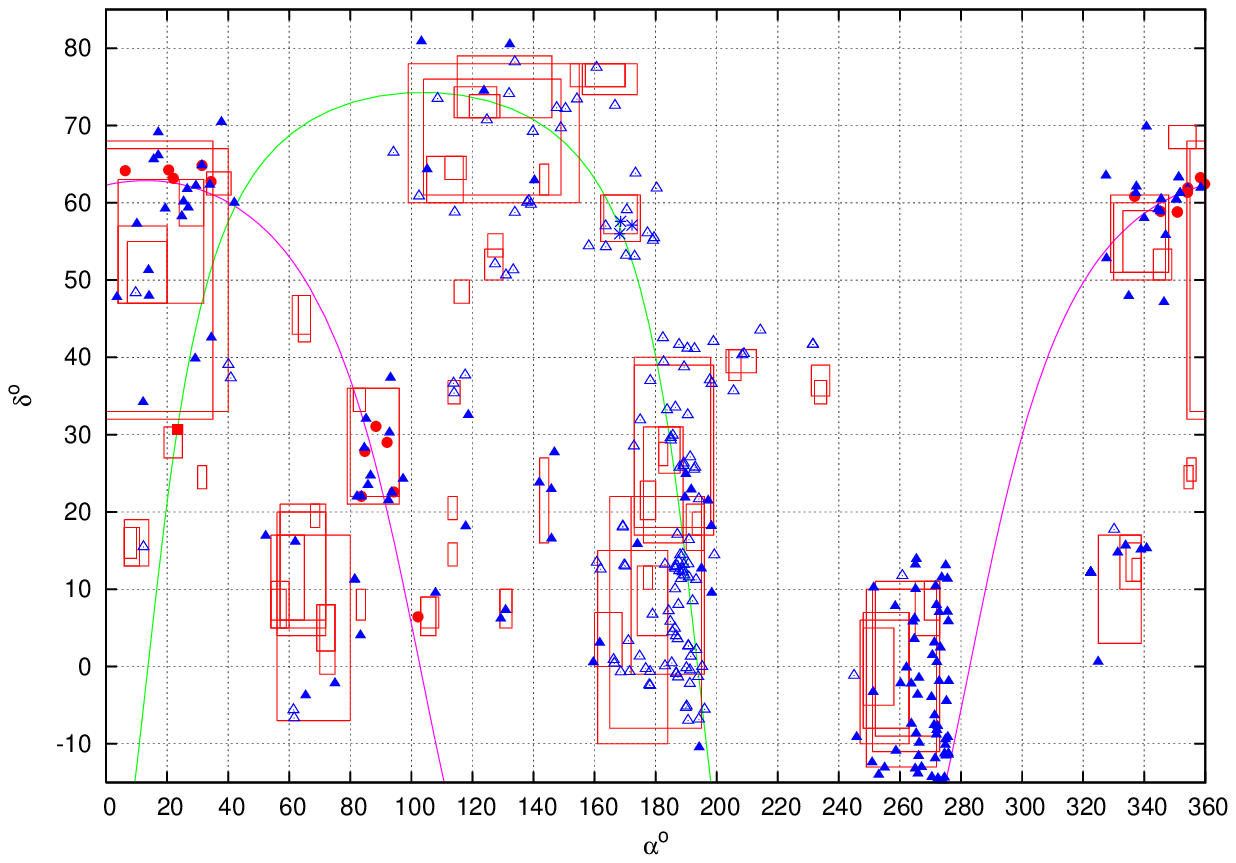}}

   Let us now consider correlations between positions of the ZEFs
   and coordinates of possible astrophysical sources of cosmic rays
   with energies near the knee.
   Figure~2 shows A-zones that have close counterparts among the ZEFs
   presented in~[\cref{we1}, \cref{we2}] and coordinates of galactic
   supernova remnants (SNRs)~\cite{Green}, pulsars~\cite{ATNF}
   (as of the state on June~19, 2006), and active galactic nuclei (AGN)
   and interacting galaxies at red shifts $z\le0.01$~\cite{Simbad}.
   All shown astrophysical objects have coordinates at angular
   distances~$\le3^\circ$ from the closest A-zone.
   This criterion selected 18 of 110 SNRs at $\delta>-16^\circ$
   with 15 of them lying inside A-zones and two of them lying
   at angular distances~$\le1.5^\circ$,
   154 of 694 pulsars, and 139 of 317 AGN and interacting galaxies.
   In particular, the selected objects include the famous SNRs
   SN1181, 3C461 (Cas~A), SN1572 (Tycho), and the Crab Nebula,
   a number of pulsars located at distances~$<0.5$~kpc from the Solar
   system, as well as the active galactic nuclei 
   NGC 3718, NGC 2681, NGC 4278, Mrk 1307, M105, M64, M65, M66, etc.,
   which have red shifts $z\le0.0033$.
   Notice that the Virgo cluster of galaxies lies inside a huge A-zone 
   located along the Supergalactic plane 
   ($\alpha\approx170^\circ\dots200^\circ$).
   It is evident that there are numerous coincidences between
   coordinates of possible astrophysical sources of cosmic rays
   with energies near the knee and positions of zones located
   in the vicinity of the Galactic and Supergalactic planes.

\asection{Acknowledgments}
   Only free, open source software was used for the investigation.
   In particular, all calculations were performed with
   GNU Octave~\cite{Octave} running in Slackware Linux.
   The research was partially supported by the Federal Program
   ``Priority Investigations in Science and Technology'' for
   2002--2006, contract No.~02.452.11.7053, and the Russian
   Foundation for Fundamental Research grant No.~05-02-16401.

\listrefs
\bye